\documentclass[twocolumn,amssymb,amsmath,amsfonts,aps,superscriptaddress,footinbib,byrevtex]{revtex4}

\usepackage{dcolumn}
\usepackage{bm}
\usepackage{graphicx}
\usepackage{times}

\begin{document}
 
\title{Does avian magnetoreception rely on both magnetite and maghemite? }

\author{Michael Winklhofer}
\affiliation{Department of Earth and Environmental Siences, University of
Munich, D-80333 M\"unchen, Germany}
%\surname{NAWI-08-0022}
 
\author{Joseph L. Kirschvink}

\affiliation{
Division of Geological and Planetary Sciences, 
California Institute of Technology, Pasadena CA 91125, USA}

%\keywords{Magnetoreception, magnetite, maghemite, ferrihydrite, homing pigeons, X-ray absorption spectroscopy, selected-area electron diffraction}
 
\begin{abstract}

Recently, a putative magnetoreceptor in the upper-beak skin of homing pigeons was
chemically characterized using X-ray flourescence and absorbtion studies 
[Fleissner \textit{et al.}, 2007. A novel concept of Fe-mineral based magnetoreception: 
histological and physiochemical data from the upper beak of homing pigeons. 
Naturwissenschaften 94 (8), 631-642]. 
In this short communication we point out that the novel concept propagated is based on conclusions 
that are not supported by data and, in fact, stand in contradiction 
to previously published crystallographic and magnetic data. 

\end{abstract}

\maketitle

\section{Introduction}\label{Intro}

Fleissner and coworkers \cite{Fleissner:07} report on their recent approach to elucidate the nature of the putative magnetoreceptor in homing pigeons. The experimental work was done very carefully and undoubtedly defines a significant progress in the field, with highlights being the reconstruction of the spatial distribution and orientation of the 
dendritic fields hosting the putative magnetoreceptor cells and the quantification of the amount of iron in these fields
using X-ray fluorscence analysis with a spatial resolution of 15~$\mu$m. Unfortunately, the experimental data presented in the paper do not justify either of the two main conclusions, namely that both magnetite and maghemite are needed for Fe-mineral based magnetoreception, and, more seriously, that the $\mu$m-sized electron-opaque platelets are monocrystals of maghemite. These problems are as follows:

\section{Non-uniqueness of XANES data}\label{XANES}

In \cite{Fleissner:07}, X-ray near-edge absorption spectroscopy (XANES) was used as a tool to 
chemically characterize the iron compounds in the dendritic field. The obtained XANES data were 
fitted to reference spectra of magnetite ($\gamma-\mathrm{Fe}_{9/3}\mathrm{O}_4$), maghemite ($\gamma-\mathrm{Fe}_{8/3}\mathrm{O}_4$, 
the fully oxidized, slightly less magnetic sister mineral to magnetite), 
and hematite ($\alpha-\mathrm{Fe}_{8/3}\mathrm{O}_4$).     
On the basis of the fitted spectra, Fleissner {\em et al} \cite{Fleissner:07} concluded that
they have unambiguously identified maghemite as the predominating iron mineral in the
dendritic fields, which is at least five times, if not ten times more abundant than magnetite. 

This conclusion is not justified because the XANES spectrum of maghemite is similar 
to those of other ferric, but non-magnetic (or only weakly magnetic) Fe-compounds not taken 
into consideration in \cite{Fleissner:07}. 
For example, goethite ($\alpha$-FeOOH), lepidocrocite ($\gamma$-FeOOH), ferrihydrite (Fe$_{10}$O$_{14}$(OH)$_2$) and even Fe(III)PO$_4$ show an equally good fit to the slope of the pigeon spectrum (Fig. 1f in \cite{Fleissner:07}) in the energy window selected in \cite{Fleissner:07} (i.e., 7116-7125 eV) as does the spectrum of maghemite (see Fig. 1 here)! Differences between these ferric iron-mineral phases become apparent only at higher energies in the spectrum (above 7125 eV), as illustrated in Fig. 1 here but not reported in \cite{Fleissner:07}. Despite these differences at higher energies, the unambiguous isolation of a particular mineral of the family of ferric oxides/hydroxides/phosphate in a mixture is difficult (see also \cite{ODay:04}). 

%\begin{widetext}
\begin{figure}
\includegraphics[width=8cm,clip=true,bb=149 420 485 709]{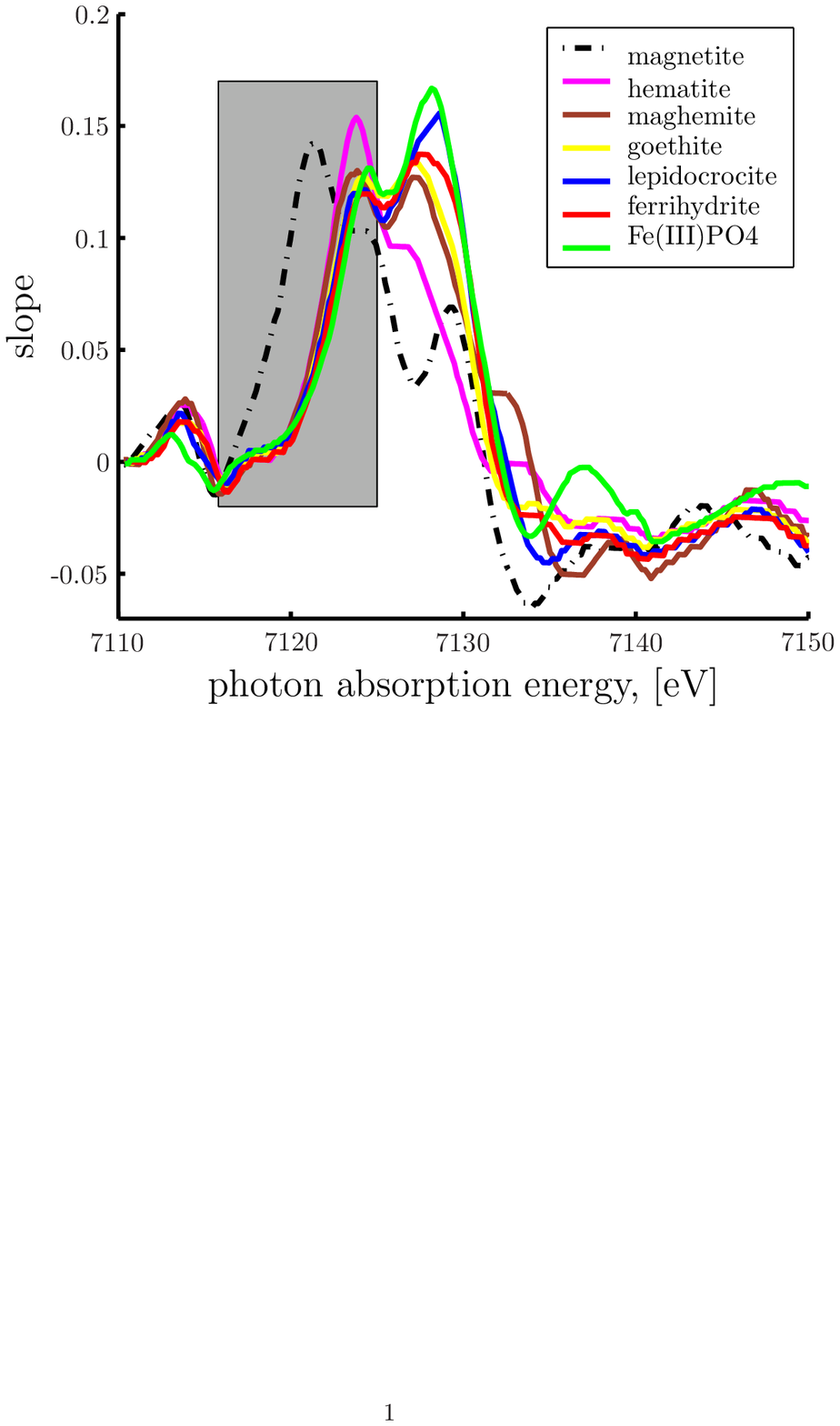}
\caption{ Differential XANES spectra (Fe-K edge) for a range of iron compounds (reference data from O'Day et al. \cite{ODay:04}). The grey patch shows the energy window selected by Fleissner {\em et al.} (\cite{Fleissner:07}, Fig 1f) to fit reference spectra of magnetite (black, dash-dotted), hematite (purple), and maghemite (brown) to the one obtained for the dendrites. 
It is obvious that it is not possible to distinguish between maghemite and other candidate ferric iron-minerals 
(goethite, lepidocrocite, ferrihydrite, ferric phosphate) within the window selected.}
\end{figure}
%\end{widetext}

For a better comparison, we therefore recommend that these other minerals be included as possible components in their curve-fitting comparisons, and that the slope in the whole energy range be fitted. In particular, the spectrum of ferrihydrite (with variable amounts of phosphorous incorporated to capture physiological ferrihydrite, see below) should be used in the curve fitting procedure, as this mineral is found ubiquitously within the cage proteins ferritin or hemosiderin that sequester iron, and it is also the mineral precursor to magnetite biomineralization in the teeth of the {\em Polyplacophoran} molluscs \cite{KirschLowen:79}.

In any case, the only conclusion that can be drawn from the XANES data is that 
pure magnetite can explain at most 20\% of the Fe compounds present in the dendritic fields.
The predominant ferric iron phase, however, cannot be unambiguously determined in the energy range
selected and the fact that only maghemite is strongly magnetic does not imply its presence.
At this stage, other candidate phases are equally likely. 

\section{No evidence for crystalline nature of the platelets}\label{SAED}

The authors go so far as to claim that the $\mu$m sized electron-opaque platelets 
arranged in chains or bands in the dendritic fields (Figs.~2c-f in \cite{Fleissner:07}) 
are monocrystals (crystallographic single domains) of maghemite. 
However, an electron-opaque region need not be crystalline. In particular, the 
size of an electron-opaque region need not represent the crystallographic domain size.
Selected-area electron diffraction (SAED) analysis would have 
been the method of choice to demonstrate the single-crystal nature of the platelets 
and provide the lattice parameter and space group. For, \bf single maghemite crystals
of that particle size and shape (1 $\mu$m x 1 $\mu$m x 0.1 $\mu$m) will produce clear 
and unmistakable, mineral-specific diffraction patterns \footnote{Although maghemite 
and magnetite have almost the same $d$ spacings, which makes their discrimination by 
means of diffraction difficult, the vacancies in maghemite are often ordered, which 
gives rise to a tetragonal superstructure, by which maghemite can be identified.}. \rm  
Yet, the authors do not support their claim by showing SAED patterns of the structures 
in question, despite their use of numerous TEM images from which such patterns would 
have been trivial to obtain.

Showing the corresponding diffractogram would have been even more important in light 
of the previous paper \cite{Fleissner:2003}, in which the electron-opaque platelets were 
reported to be amorphous (i.e., they showed no diffraction rings in SAED) and to be 
associated with high concentrations of iron and phosphorous. 
Those characteristics are more typical of physiological ferrihydrite, which also contains 
variable amounts of phosphorous but yields characteristically poor diffraction patterns. 
A similar mistake - misidentifying concentrations of ferrihydrite for magnetite/maghemite 
- was made about 10 years ago in the study of iron in the abdomen of honeybees \cite{HsuLi:94}; 
see discussions by \cite{NicholLocke:95,KirschvinkWalker:95,Nesson:95}. 

The conclusion that the platelets are monocrystals of maghemite is also at odds with
bulk magnetic measurements \cite{Hanzlik:2000}, which showed that the predominant magnetic
phase is superparamagnetic. Judging from the high abundance of platelets relative to
clusters in the dendritic fields, we would expect to see a pronounced and stable magnetic
remenance (at room temperature) if each platelet (dimensions 1 $\mu$m x 1 $\mu$m x 0.1  $\mu$m) 
were to represent a single-crystal of maghemite. This is not observed.

\section{Why the distinction between magnetite and maghemite is not decisive}\label{Sol3}

Unlike conveyed by \cite{Fleissner:07}, magnetite and maghemite actually are closely
related to one another, both structurally and magnetically (see next Section). 
From the point of view of crystal chemistry, they are both inverse 
spinels \footnote{An inverse spinel has (B)[AB]O$_4$ structure, where the tetrahedrally
coordinated sites (indicated by round brackets) are occupied by cation B, and the octahedral 
sites (indicated by square brackets) are occupied by cations A and B. To emphasize the inverse 
spinel structure, the formula units of magnetite and maghemite can also be written as $\mathrm{Fe}^{3+}[\mathrm{Fe}^{2+}\mathrm{Fe}^{3+}]\mathrm{O}^{2-}_4$ and
$\mathrm{Fe}^{3+}[\mathrm{Fe}^{3+}_{5/3}\square_{1/3}]\mathrm{O}^{2-}_4$, respectively. 
}, have similar lattice parameters, and differ only in the oxidation state of iron,
with maghemite representing fully oxidized,  cation-deficient magnetite, 
where every ninth Fe cation is substituted by a vacancy to 
compensate for the charge imbalance. This is why there is a solid-solution 
series between the two end members, $\mathrm{Fe}^{3+}_{2+2 z/3}\mathrm{Fe}^{2+}_{1 - z}\square{}_{z/3}\mathrm{O}^{2-}_4$, where $z$ is the oxidation parameter ($0 \leq z \leq 1 $) and $\square$ denotes a vacancy, see for example \cite{ReadmanOReilly:72,Lindsley:81}.  

Note that even magnetotactic bacteria (MB) - the model organisms for 
magnetite biomineralization - 
do not contain 100\% end-member magnetite as can easily be seen by the fact that 
MB have lower Verwey transition temperatures ($T_\mathrm{v} \sim  105 -115$~K)\cite{Moskowitz:93a,Weiss:2004,pan:05b} than ideal stoichiometric magnetite has ($T_\mathrm{v}=125$~K) \cite{Walz:2002}. Since $T_\mathrm{v}$ decreases with increasing cation deficiency (maghemitization) in magnetite \cite{Aragon:85}, MB contain partially maghemitized magnetite crystals, which reflects the oxidation state of
the local environment in which they are grown \cite{Kopp:07}. 
Nonetheless we refer to them as magnetite, without making a big deal 
out of the oxidation state.

\section{Why iron-based magnetoreception can perform with one magnetic Fe-mineral}\label{Femineral}

In their abstract, Fleissner {\em et al} \cite{Fleissner:07} state, 
"in this paper we show that iron-based magnetoreception needs the presence of both of these iron minerals, $\ldots$",
Again, we find no support for this statement in the published work. In essence, the authors 
claim that magnetite and maghemite have distinctly different magnetic properties. For example, 
it is written in \cite{Fleissner:07} that 
"The flat quadratic shape [of the electron-opaque platelets] perfectly fits the 
magnetocrystalline anisotropy of maghemite that coincides with the direction of the cubic axis". 
That statement is wrong for two reasons. For one, the postulated quadratic shape of the electron-opaque 
platelets cannot be taken as crystallographic evidence of a (100) facet, let alone as evidence of 
maghemite (see Section~\ref{SAED}). Secondly, \bf both magnetite and magnetite have a negative cubic 
magnetocrystalline anisotropy constant \rm $K_1$, 
which means that the preferential axes of the magnetocrystalline anisotropy are the $\langle 111 \rangle$ axes 
(i.e., normal to the faces of an octahedron) and not the crystallographic $\langle 100 \rangle$ axes 
(normal to the faces of a cube) (e.g., \cite{Stacey:74}). This is also well-known 
from paleomagnetic studies, where magnetite is often found to be oxidized, but still carries 
the originally recorded remanence direction. If the sign of $K_1$ were to change during oxidation, 
the remanence direction would change too, which is not observed.

\bf Whether the magnetic core in the receptor is made up of magnetite or maghemite
is not a decisive physiological parameter. \rm The single most important material 
parameter for Fe-mineral based magnetoreception is the saturation magnetization, 
$M_{\mathrm{s}}$, which defines the sensitivity of the receptor to detect minute 
changes in geomagnetic field strength. The value of $M_{\mathrm{s}}$ is 480~G for 
pure magnetite compared to 380~G for pure maghemite (at room temperature), that is, 
both values are significantly larger than the geomagnetic field strength (0.25-0.6~G) and 
thus can locally amplify the external field by two orders of magnitude.

While the sign of $K_1$ for maghemite is negative, different values of $K_1$ have been
reported in the literature, ranging from one-third to 80\% the value of magnetite \cite{Dunlop:97,Stacey:74}.
Probably, these differences reflect different degrees of vacancy ordering in the maghemite
crystals studied. Nevertheless, even with one-third the value of $K_1$ for magnetite, maghemite 
is magnetically not much softer than magnetite. Importantly, the value of $K_1$ is not directly 
related to magnetic hardness or softness, but rather determines the ease with which a desired 
hardness (or softness) can be achieved. Coercivity, an actual measure of magnetic hardness, is 
determined mainly by additional structural features such as lattice defects, grain boundaries, 
magnetic particle size and shape \cite{Hubert:98}. The platelets in question (if they were to 
represent defect-free single crystals of magnetite or of magnetite) would behave as soft magnets 
(easily magnetizeable) 
primarily because of their dimensions and the resulting shape anisotropy, which overrides the magnetocrystalline
anisotropy. We conducted micromagnetic calculations for such platelets and found that a 
magnetite composition yields a slightly higher susceptibility (i.e., is more easily magnetizeable)
than a maghemite composition does, even though maghemite was modelled with a $K_1$ of one-third 
the value of $K_1$ for magnetite. Thus, the higher $M_{\mathrm{s}}$ value of magnetite compared 
to maghemite more than outweighs the intrinsically harder character of magnetite and a higher 
magnetic-field amplification can be realized on the basis of magnetite.

Therefore, the "novel concept of Fe-mineral-based magnetoreception" 
\cite{Fleissner:07,Fleissner:07b} first and foremost relies on a bimodal distribution of magnetic 
particle-sizes and, with it, of domain states (superparamanetic, 
SP vs mulit-domain, MD), but not 
on the presence of two Fe oxides which have very similar magnetic properties 
anyway\footnote{Note that despite their similar magnetic properties,
magnetite and maghemite have different electronic properties. 
Maghemite is an insolator, while magnetite is a half-metal,
that is, electronic conductivity is spin-polarized and 
depends on the magnetization direction. Whether a magnetite-based
magnetoreceptor can exploit these intriguing electronic properties 
is a different question. One way of realizing a magnetically gated 
biological electrical circuit with magnetite was suggested in 
\cite{KirschGould:81}, referred to as "the membrane shortcut model".}. In essence, what has been suggested 
in \cite{Fleissner:07} and followed up on in \cite{Fleissner:07b,Solovjov:07,Solovjov:07b} 
is a hybrid-magnetoreceptor 
(combining adjacent magnetic elements with different properties, here SP and MD),
conceptually similar to the ones that have already been discussed 
in the literature (e.g. \cite{ShcherbakovWinklhofer:99}, there SP and SD).\\

\section{Conclusions} 
 
The experimental data presented in \cite{Fleissner:07} do not justify the 
reinterpretation of the crystallographic results reported in \cite{Fleissner:2003},
where the $\mu$m-sized electron-opaque platelets were found to be amorphous. The 
XANES spectra cannot be uniquely attributed to maghemite, but may as well be 
explained by physiological ferrihydrite, which is only weakly magnetic, but
due to its poorly crystalline (amorphous) nature and high concentration of 
iron and phosphorous consistent with the results previously reported in
\cite{Fleissner:2003}. 
The claim that both magnetite and maghemite are needed to realize a biological
magnetometer is based on the wrong assumption that maghemite and magnetite
are positive-anisotropy and negative-anisotropy cubic materials, respectively.
However, they are both negative-anisotropy cubic materials ($K_1<0$) and rather
similar in their magnetic properties. 

\subsection*{Acknowledgments} 

We acknowledge funding from the Human Frontier Science Program Organization, HFSP 
research grant RGP28/2007.

\subsection*{Note} 

\em This is an extended version of a comment 
submitted to {\em Naturwissenschaften} on 17 June 2007. 
A manuscript number was eventually assigned on 23 Jan 2008 
(NAWI-08-0022) and the manuscript has since been under review. 
The complementary diffraction data requested have not been
furnished yet.\rm

\section{References}
\bibliographystyle{apsrev}
\bibliography{myJrnlabbrv,natwisscom}

\begin{thebibliography}{25}
\expandafter\ifx\csname natexlab\endcsname\relax\def\natexlab#1{#1}\fi
\expandafter\ifx\csname bibnamefont\endcsname\relax
  \def\bibnamefont#1{#1}\fi
\expandafter\ifx\csname bibfnamefont\endcsname\relax
  \def\bibfnamefont#1{#1}\fi
\expandafter\ifx\csname citenamefont\endcsname\relax
  \def\citenamefont#1{#1}\fi
\expandafter\ifx\csname url\endcsname\relax
  \def\url#1{\texttt{#1}}\fi
\expandafter\ifx\csname urlprefix\endcsname\relax\def\urlprefix{URL }\fi
\providecommand{\bibinfo}[2]{#2}
\providecommand{\eprint}[2][]{\url{#2}}

\bibitem[{\citenamefont{Fleissner
  et~al.}(2007{\natexlab{a}})\citenamefont{Fleissner, Stahl, Thalau,
  Falkenberg, and Fleissner}}]{Fleissner:07}
\bibinfo{author}{\bibfnamefont{G.}~\bibnamefont{Fleissner}},
  \bibinfo{author}{\bibfnamefont{B.}~\bibnamefont{Stahl}},
  \bibinfo{author}{\bibfnamefont{P.}~\bibnamefont{Thalau}},
  \bibinfo{author}{\bibfnamefont{G.}~\bibnamefont{Falkenberg}},
  \bibnamefont{and}
  \bibinfo{author}{\bibfnamefont{G.}~\bibnamefont{Fleissner}},
  \bibinfo{journal}{Naturwissenschaften} \textbf{\bibinfo{volume}{94}},
  \bibinfo{pages}{631} (\bibinfo{year}{2007}{\natexlab{a}}).

\bibitem[{\citenamefont{{O'Day} et~al.}(2004)\citenamefont{{O'Day}, Rivera,
  Root, and Carroll}}]{ODay:04}
\bibinfo{author}{\bibfnamefont{P.}~\bibnamefont{{O'Day}}},
  \bibinfo{author}{\bibfnamefont{N.}~\bibnamefont{Rivera}},
  \bibinfo{author}{\bibfnamefont{R.}~\bibnamefont{Root}}, \bibnamefont{and}
  \bibinfo{author}{\bibfnamefont{S.~A.} \bibnamefont{Carroll}},
  \bibinfo{journal}{Am. Mineral.} \textbf{\bibinfo{volume}{89}},
  \bibinfo{pages}{572} (\bibinfo{year}{2004}).

\bibitem[{\citenamefont{Kirschvink and Lowenstam}(1979)}]{KirschLowen:79}
\bibinfo{author}{\bibfnamefont{J.~L.} \bibnamefont{Kirschvink}}
  \bibnamefont{and} \bibinfo{author}{\bibfnamefont{H.~A.}
  \bibnamefont{Lowenstam}}, \bibinfo{journal}{Earth Planet. Sci. Lett.}
  \textbf{\bibinfo{volume}{4}}, \bibinfo{pages}{193} (\bibinfo{year}{1979}).

\bibitem[{\citenamefont{Fleissner et~al.}(2003)\citenamefont{Fleissner,
  Holtkamp-R\"otzler, Hanzlik, Winklhofer, Fleissner, Petersen, and
  Wiltschko}}]{Fleissner:2003}
\bibinfo{author}{\bibfnamefont{G.}~\bibnamefont{Fleissner}},
  \bibinfo{author}{\bibfnamefont{E.}~\bibnamefont{Holtkamp-R\"otzler}},
  \bibinfo{author}{\bibfnamefont{M.}~\bibnamefont{Hanzlik}},
  \bibinfo{author}{\bibfnamefont{M.}~\bibnamefont{Winklhofer}},
  \bibinfo{author}{\bibfnamefont{G.}~\bibnamefont{Fleissner}},
  \bibinfo{author}{\bibfnamefont{N.}~\bibnamefont{Petersen}}, \bibnamefont{and}
  \bibinfo{author}{\bibfnamefont{W.}~\bibnamefont{Wiltschko}},
  \bibinfo{journal}{J. Comp. Neurol.} \textbf{\bibinfo{volume}{458}},
  \bibinfo{pages}{350} (\bibinfo{year}{2003}).

\bibitem[{\citenamefont{Hsu and Li}(1994)}]{HsuLi:94}
\bibinfo{author}{\bibfnamefont{C.-Y.} \bibnamefont{Hsu}} \bibnamefont{and}
  \bibinfo{author}{\bibfnamefont{C.-W.} \bibnamefont{Li}},
  \bibinfo{journal}{Science} \textbf{\bibinfo{volume}{265}},
  \bibinfo{pages}{95} (\bibinfo{year}{1994}).

\bibitem[{\citenamefont{Nichol and Locke}(1995)}]{NicholLocke:95}
\bibinfo{author}{\bibfnamefont{H.}~\bibnamefont{Nichol}} \bibnamefont{and}
  \bibinfo{author}{\bibfnamefont{M.}~\bibnamefont{Locke}},
  \bibinfo{journal}{Science} \textbf{\bibinfo{volume}{269}},
  \bibinfo{pages}{1888} (\bibinfo{year}{1995}).

\bibitem[{\citenamefont{Kirschvink and Walker}(1995)}]{KirschvinkWalker:95}
\bibinfo{author}{\bibfnamefont{J.~L.} \bibnamefont{Kirschvink}}
  \bibnamefont{and} \bibinfo{author}{\bibfnamefont{M.~M.}
  \bibnamefont{Walker}}, \bibinfo{journal}{Science}
  \textbf{\bibinfo{volume}{269}}, \bibinfo{pages}{1889} (\bibinfo{year}{1995}).

\bibitem[{\citenamefont{Nesson}(1995)}]{Nesson:95}
\bibinfo{author}{\bibfnamefont{M.~H.} \bibnamefont{Nesson}},
  \bibinfo{journal}{Science} \textbf{\bibinfo{volume}{269}},
  \bibinfo{pages}{1889} (\bibinfo{year}{1995}).

\bibitem[{\citenamefont{Hanzlik et~al.}(2000)\citenamefont{Hanzlik, Heunemann,
  Holtkamp-R\"otzler, Winklhofer, Petersen, and Fleissner}}]{Hanzlik:2000}
\bibinfo{author}{\bibfnamefont{M.}~\bibnamefont{Hanzlik}},
  \bibinfo{author}{\bibfnamefont{C.}~\bibnamefont{Heunemann}},
  \bibinfo{author}{\bibfnamefont{E.}~\bibnamefont{Holtkamp-R\"otzler}},
  \bibinfo{author}{\bibfnamefont{M.}~\bibnamefont{Winklhofer}},
  \bibinfo{author}{\bibfnamefont{N.}~\bibnamefont{Petersen}}, \bibnamefont{and}
  \bibinfo{author}{\bibfnamefont{G.}~\bibnamefont{Fleissner}},
  \bibinfo{journal}{Biometals} \textbf{\bibinfo{volume}{13}},
  \bibinfo{pages}{325} (\bibinfo{year}{2000}).

\bibitem[{\citenamefont{Readman and {O'Reilly}}(1972)}]{ReadmanOReilly:72}
\bibinfo{author}{\bibfnamefont{P.~W.} \bibnamefont{Readman}} \bibnamefont{and}
  \bibinfo{author}{\bibfnamefont{W.}~\bibnamefont{{O'Reilly}}},
  \bibinfo{journal}{J. Geomagn. Geoelec.} \textbf{\bibinfo{volume}{24}},
  \bibinfo{pages}{69} (\bibinfo{year}{1972}).

\bibitem[{\citenamefont{Lindsley}(1981)}]{Lindsley:81}
\bibinfo{author}{\bibfnamefont{D.~H.} \bibnamefont{Lindsley}}, in
  \emph{\bibinfo{booktitle}{{Reviews in Mineralogy, vol 3: Oxide Minerals}}}
  (\bibinfo{publisher}{{Mineralogical Society of America}},
  \bibinfo{year}{1981}), pp. \bibinfo{pages}{L1--L60}.

\bibitem[{\citenamefont{Moskowitz et~al.}(1993)\citenamefont{Moskowitz, Bruce,
  Frankel, and Bazylinski}}]{Moskowitz:93a}
\bibinfo{author}{\bibfnamefont{B.}~\bibnamefont{Moskowitz}},
  \bibinfo{author}{\bibfnamefont{M.}~\bibnamefont{Bruce}},
  \bibinfo{author}{\bibfnamefont{R.}~\bibnamefont{Frankel}}, \bibnamefont{and}
  \bibinfo{author}{\bibfnamefont{D.}~\bibnamefont{Bazylinski}},
  \bibinfo{journal}{Earth Planet. Sci. Lett.} \textbf{\bibinfo{volume}{120}},
  \bibinfo{pages}{283} (\bibinfo{year}{1993}).

\bibitem[{\citenamefont{Weiss et~al.}(2002)\citenamefont{Weiss, Kim,
  Kirschvink, Kopp, Sankaran, Kobayashi, and Komeili}}]{Weiss:2004}
\bibinfo{author}{\bibfnamefont{B.~P.} \bibnamefont{Weiss}},
  \bibinfo{author}{\bibfnamefont{S.~S.} \bibnamefont{Kim}},
  \bibinfo{author}{\bibfnamefont{J.~L.} \bibnamefont{Kirschvink}},
  \bibinfo{author}{\bibfnamefont{R.~E.} \bibnamefont{Kopp}},
  \bibinfo{author}{\bibfnamefont{M.}~\bibnamefont{Sankaran}},
  \bibinfo{author}{\bibfnamefont{A.}~\bibnamefont{Kobayashi}},
  \bibnamefont{and} \bibinfo{author}{\bibfnamefont{A.}~\bibnamefont{Komeili}},
  \bibinfo{journal}{Earth Planet. Sci. Lett.} \textbf{\bibinfo{volume}{224}},
  \bibinfo{pages}{73} (\bibinfo{year}{2002}).

\bibitem[{\citenamefont{Pan et~al.}(2005)\citenamefont{Pan, Petersen,
  Winklhofer, Davila, Liu, Frederichs, Hanzlik, and Zhu}}]{pan:05b}
\bibinfo{author}{\bibfnamefont{Y.~X.} \bibnamefont{Pan}},
  \bibinfo{author}{\bibfnamefont{N.}~\bibnamefont{Petersen}},
  \bibinfo{author}{\bibfnamefont{M.}~\bibnamefont{Winklhofer}},
  \bibinfo{author}{\bibfnamefont{A.~F.} \bibnamefont{Davila}},
  \bibinfo{author}{\bibfnamefont{Q.~S.} \bibnamefont{Liu}},
  \bibinfo{author}{\bibfnamefont{T.}~\bibnamefont{Frederichs}},
  \bibinfo{author}{\bibfnamefont{M.}~\bibnamefont{Hanzlik}}, \bibnamefont{and}
  \bibinfo{author}{\bibfnamefont{R.~X.} \bibnamefont{Zhu}},
  \bibinfo{journal}{Earth Planet Sci. Lett} \textbf{\bibinfo{volume}{237}},
  \bibinfo{pages}{311} (\bibinfo{year}{2005}).

\bibitem[{\citenamefont{Walz}(2002)}]{Walz:2002}
\bibinfo{author}{\bibfnamefont{F.}~\bibnamefont{Walz}}, \bibinfo{journal}{J.
  Phys. Cond. Matt.} \textbf{\bibinfo{volume}{14}}, \bibinfo{pages}{R285}
  (\bibinfo{year}{2002}).

\bibitem[{\citenamefont{Aragon et~al.}(1985)\citenamefont{Aragon, Buttrey,
  Shepherd, and Honig}}]{Aragon:85}
\bibinfo{author}{\bibfnamefont{R.}~\bibnamefont{Aragon}},
  \bibinfo{author}{\bibfnamefont{D.~J.} \bibnamefont{Buttrey}},
  \bibinfo{author}{\bibfnamefont{J.~P.} \bibnamefont{Shepherd}},
  \bibnamefont{and} \bibinfo{author}{\bibfnamefont{J.~M.} \bibnamefont{Honig}},
  \bibinfo{journal}{Phys Rev. B.} \textbf{\bibinfo{volume}{31}},
  \bibinfo{pages}{430} (\bibinfo{year}{1985}).

\bibitem[{\citenamefont{Kopp}(2007)}]{Kopp:07}
\bibinfo{author}{\bibfnamefont{R.~E.} \bibnamefont{Kopp}},
  \emph{\bibinfo{title}{The identification and interpretation of microbial
  geobiomagnetism}} (\bibinfo{publisher}{PhD Thesis, California Institute of
  Technology, on-line at
  http://resolver.caltech.edu/CaltechETD:etd-04122007-135320},
  \bibinfo{address}{Pasadena}, \bibinfo{year}{2007}).

\bibitem[{\citenamefont{Stacey and Banerjee}(1974)}]{Stacey:74}
\bibinfo{author}{\bibfnamefont{F.~D.} \bibnamefont{Stacey}} \bibnamefont{and}
  \bibinfo{author}{\bibfnamefont{S.~K.} \bibnamefont{Banerjee}},
  \emph{\bibinfo{title}{The physical principles of rock magnetism}}
  (\bibinfo{publisher}{Elsevier}, \bibinfo{address}{Amsterdam},
  \bibinfo{year}{1974}).

\bibitem[{\citenamefont{Dunlop and \"Ozdemir}(1997)}]{Dunlop:97}
\bibinfo{author}{\bibfnamefont{D.~J.} \bibnamefont{Dunlop}} \bibnamefont{and}
  \bibinfo{author}{\bibfnamefont{O.}~\bibnamefont{\"Ozdemir}},
  \emph{\bibinfo{title}{Rock Magnetism: Fundamentals and Frontiers}}
  (\bibinfo{publisher}{Cambridge University Press}, \bibinfo{year}{1997}).

\bibitem[{\citenamefont{Hubert and Sch\"afer}(1998)}]{Hubert:98}
\bibinfo{author}{\bibfnamefont{A.}~\bibnamefont{Hubert}} \bibnamefont{and}
  \bibinfo{author}{\bibfnamefont{R.}~\bibnamefont{Sch\"afer}},
  \emph{\bibinfo{title}{Magnetic domains}} (\bibinfo{publisher}{Springer},
  \bibinfo{address}{Berlin}, \bibinfo{year}{1998}).

\bibitem[{\citenamefont{Fleissner
  et~al.}(2007{\natexlab{b}})\citenamefont{Fleissner, Stahl, Thalau,
  Falkenberg, and Fleissner}}]{Fleissner:07b}
\bibinfo{author}{\bibfnamefont{G.}~\bibnamefont{Fleissner}},
  \bibinfo{author}{\bibfnamefont{B.}~\bibnamefont{Stahl}},
  \bibinfo{author}{\bibfnamefont{P.}~\bibnamefont{Thalau}},
  \bibinfo{author}{\bibfnamefont{G.}~\bibnamefont{Falkenberg}},
  \bibnamefont{and}
  \bibinfo{author}{\bibfnamefont{G.}~\bibnamefont{Fleissner}},
  \bibinfo{journal}{J. Ornithol.} \textbf{\bibinfo{volume}{94}},
  \bibinfo{pages}{S643} (\bibinfo{year}{2007}{\natexlab{b}}).

\bibitem[{\citenamefont{Solvo'yov and
  Greiner}(2007{\natexlab{a}})}]{Solovjov:07}
\bibinfo{author}{\bibfnamefont{I.~A.} \bibnamefont{Solvo'yov}}
  \bibnamefont{and} \bibinfo{author}{\bibfnamefont{W.}~\bibnamefont{Greiner}},
  \bibinfo{journal}{Biophys. J.} \textbf{\bibinfo{volume}{93}},
  \bibinfo{pages}{1493} (\bibinfo{year}{2007}{\natexlab{a}}).

\bibitem[{\citenamefont{Solvo'yov and
  Greiner}(2007{\natexlab{b}})}]{Solovjov:07b}
\bibinfo{author}{\bibfnamefont{I.~A.} \bibnamefont{Solvo'yov}}
  \bibnamefont{and} \bibinfo{author}{\bibfnamefont{W.}~\bibnamefont{Greiner}},
  \emph{\bibinfo{title}{Towards understanding of birds magnetoreceptor
  mechanism}} (\bibinfo{year}{2007}{\natexlab{b}}),
  \urlprefix\url{http://arxiv.org/abs/0704.1763}.

\bibitem[{\citenamefont{Shcherbakov and
  Winklhofer}(1999)}]{ShcherbakovWinklhofer:99}
\bibinfo{author}{\bibfnamefont{V.~P.} \bibnamefont{Shcherbakov}}
  \bibnamefont{and}
  \bibinfo{author}{\bibfnamefont{M.}~\bibnamefont{Winklhofer}},
  \bibinfo{journal}{Eur. Biophys. J.} \textbf{\bibinfo{volume}{28}},
  \bibinfo{pages}{380} (\bibinfo{year}{1999}).

\bibitem[{\citenamefont{Kirschvink and Gould}(1981)}]{KirschGould:81}
\bibinfo{author}{\bibfnamefont{J.~L.} \bibnamefont{Kirschvink}}
  \bibnamefont{and} \bibinfo{author}{\bibfnamefont{J.~L.} \bibnamefont{Gould}},
  \bibinfo{journal}{BioSystems} \textbf{\bibinfo{volume}{13}},
  \bibinfo{pages}{181} (\bibinfo{year}{1981}).

\end{thebibliography}

 \end{document}